\documentclass[prb,reprint,superscriptaddress,floatfix]{revtex4-1}
\usepackage[utf8]{inputenc}
\usepackage{textcomp}
\usepackage[T1]{fontenc}
\usepackage{amsmath}
\usepackage{amssymb}
\usepackage{setspace}
\usepackage{physics}	
\usepackage{comment}
\usepackage{float}
\usepackage{bm}
\usepackage{graphicx}
\graphicspath{./}
\usepackage{placeins}
\usepackage{hyperref}
\hypersetup{
 unicode=false,			
 pdftoolbar=true,		
 pdffitwindow=false,		
 pdfstartview={FitH},		
 pdfcreator={pdflatex},		
 pdfproducer={vim},		
 pdfnewwindow=true,		
 colorlinks=true,		
 linktoc=page,			
 linkcolor=blue,		
 citecolor=blue,		
 filecolor=blue,		
 urlcolor=blue			
}

\newcommand{\up}{{\uparrow}}
\newcommand{\down}{{\downarrow}}
\newcommand{\hc}{\text{h.c.}}

\newcommand{\Lleg}{\mathrm{L}}
\newcommand{\Rleg}{\mathrm{R}}

\begin{document}

\title{Shuttling of $\mathbb{Z}_4$ parafermions in an electronic ladder model}

\date{\today}

\author{Botond Osv\'ath}
\affiliation{
Department of Physics of Complex Systems, ELTE E\"otv\"os Lor\'and University, H-1117, Budapest,
Hungary}
\author{Gergely Barcza}
\email{barcza.gergely@wigner.hu}
\affiliation{Wigner Research Centre for Physics, H-1525, Budapest, Hungary}
\author{L\'aszl\'o Oroszl\'any}
\email{laszlo.oroszlany@ttk.elte.hu}
\affiliation{
Department of Physics of Complex Systems, ELTE E\"otv\"os Lor\'and University, H-1117, Budapest,
Hungary}
\affiliation{Wigner Research Centre for Physics, H-1525, Budapest, Hungary}

\begin{abstract}
Parafermions with non-Abelian statistics have been proposed as a promising platform for quantum computation, potentially enabling a broader set of topologically protected gates than Majorana fermions. 
The experimental and theoretical exploration of these exotic quasiparticles remains challenging, as their stability is linked to strong electron-electron interactions. 
A key step toward practical applications is the controlled shuttling of parafermionic modes, which is required for implementing geometric braiding operations.
In the present work, we investigate the real-time dynamics of the elementary shuttling process by applying a combination of the density matrix renormalization group and the time-dependent variational principle approaches. 
We analyze the transport of $\mathbb{Z}_4$ parafermion edge states and assess the corresponding adiabatic speed limit under experimentally relevant conditions. 
\end{abstract}

\maketitle

\section{Introduction} 
The pursuit of fault-tolerant quantum computing hinges on the development of qubits that are inherently robust against decoherence and local noise.
Topological quantum computing offers a hardware-level solution to this challenge by encoding quantum information in non-local degrees of freedom of exotic anyons, making it intrinsically resilient to local perturbations.
Although significant attention has been directed toward Majorana zero modes (MZMs) -- which correspond to a $\mathbb{Z}_2$ topological index -- their non-Abelian statistics alone are insufficient for universal quantum computation~\cite{Das_Sarma_RevModPhys.80.1083,Sarma2015}.
Consequently, there is growing interest in realizing parafermions~\cite{Fendley_2014}, which generalize these excitations to a $\mathbb{Z}_n$ index (with $n > 2$) and provide a richer set of topologically protected operations necessary for universal quantum computation~\cite{Hastings_2013}. 
A related important aspect of parafermionic modes is that they might implement various logical units of information. Specifically, the $\mathbb{Z}_4$ manifold can encode a 4-level qudit whose braiding operation generates the entire Clifford group~\cite{hutter2016quantum}. In case of a two-qubit representation, the qubits might also be manipulated by non-Clifford gates implemented through mapping between Majorana fermions and $\mathbb{Z}_4$ parafermions~\cite{Safwan_2025}.
Furthermore, by applying a parity constraint, a single logical qubit can be realized which is highly resilient against quasiparticle poisoning compared to MZM-based realization as a single-electron tunneling event kicks out of the computational space rather than flipping the qubit state~\cite{hutter2016quantum}. 

MZMs can be implemented in systems where interactions are not important, such as a one-dimensional spinless $p$-wave superconductor~\cite{Kitaev_2001}.
In contrast, parafermions require, in general, strong electron-electron interactions~\cite{alicea_fendley}. 
To realize them, various proposals have been put forward relying on, e.g.,   superconductor-fractional Chern insulator hybrids~\cite{FQH_para_clarke2013exotic,FQH_para_PhysRevX.2.041002} and flat bands of moire heterostructures~\cite{Liu2025}. Furthermore, interactions can also induce parafermion modes in quantum spin Hall insulators~\cite{zhang,orth2015non,Fleckenstein2019} or in a bundle of spinless nanowires~\cite{Klinovaja_PhysRevLett.112.246403}.
Recently, a mapping between $\mathbb{Z}_4$ parafermions and non-interacting spinful fermions was studied, where the Gentile statistics of parafermionic systems can be elucidated, although these systems lack isolated parafermionic zero modes~\cite{mccann2026noninteracting}. 

Inspired by these developments, an interacting electronic ladder model capable of hosting $\mathbb{Z}_4$ parafermions was also proposed~\cite{Osvath_2024}, which can potentially be implemented with highly tunable quantum dot arrays\cite{EIGHT_QDOT_LADDER_hsiao2023exciton,TWO_DOT_KITAEV_EXP_2023realization,THREE_DOT_KITAEV_EXP_bordin2023crossed}.
Alternatively, ultracold atomic ladders \cite{CA_Ladder_li2013topological} combined with spin-orbit coupling mimicking mechanisms\cite{CA_SOC_SPIELMAN1_lin2011spin,CA_SOC2_galitski2013spin,CA_SOC3_SPIELMAN2_valdes2021topological} could also efficiently realize our model. 

With an established platform for topological excitations implementing information units, the next crucial step towards efficient quantum computations is understanding the dynamics required to implement fault-tolerant braiding protocols which realize various topologically protected quantum gates~\cite{alicea_fendley,hutter2016quantum}. 
Practical exchange operations can work only in a specific frequency window constrained by characteristic time scales~\cite{Das_Sarma_RevModPhys.80.1083,Knapp_2016,para_braiding_solofo}. In particular, the braiding must be slow enough to preserve adiabaticity of the ground state, which is primarily characterized by the energy gap of the system.
However, braiding must be fast enough to minimize the decoherence induced by finite energy splitting of the computational subspace and quasi-particle poisoning~\cite{hutter2016quantum}.
When parafermions are implemented in nanowire-based systems, the edge states can be efficiently braided using T-junction setups to avoid bringing them close together and thus splitting the critical energy degeneracy of the parafermions  while exchanging their position~\cite{Alicea2011}.
The elementary cornerstone of the physical braiding procedure is the controlled transport, or "shuttling" of the modes along the wires~\cite{alicea_fendley}.
Therefore, as an initial step towards optimizing swapping protocols, it is critical to explore the actual diabatic error profile in topological systems of interest.
For MZMs, a slow shuttling approach has been extensively studied~\cite{Bauer2018,mzm_shuttle_truong_1, lr2b-nmrk, mzm_shuttle_truong_2}, where local electric gates are used to sequentially tune sections of the wire throughout the topological phase transition.
As a result, shuttling of MZMs is found to follow exponential decay estimated by Landau-Zener theory only for shorter time scales~\cite{Bauer2018}, and power-law corrections might become dominant for sufficiently large relaxation times due to the nonanalytic behavior of the tuning function~\cite{mzm_shuttle_truong_1}.
Regarding the twofold odd and even (quasi)-degeneracy of the $\mathbb{Z}_4$ parafermion bound states, their actual diabatic error model might be even more subtle.

In this work, we investigate the finite-time shuttling of $\mathbb{Z}_4$ parafermions within a previously established strongly interacting electronic  model~\cite{Osvath_2024}. 
In the considered ladder setup, parafermionic bound states are localized at the two interfaces between an interacting central segment and superconducting terminals.
We adapt a slow transport protocol~\cite{mzm_shuttle_truong_1} by implementing a simultaneous local modulation of the interaction strength and superconducting pairing to shift one of the interfaces, and thus the localized edge states.
We employ the density matrix renormalization group~\cite{White-1992b} (DMRG) approach to characterize the static properties of the correlated system and the time-dependent variational principle~\cite{Haegaman_TDVP_PhysRevLett.107.070601} (TDVP) to simulate its quantum many-body dynamics. 
We perform a rigorous study to accurately quantify the diabatic error sources of the parafermionic subspace by marginalizing the numerical errors intrinsic to the applied computational approximations.
Our predictions give a realistic upper bound estimate of the stable shuttling frequency of the $\mathbb{Z}_4$ parafermion model tailored for quantum dot arrays.

The remainder of the paper is structured as follows. 
In Sect.~\ref{sect:methods}, the applied numerical methods are overviewed, focusing specifically on their control parameters that determine the precision of the simulations.
In Sect.~\ref{sect:modeling}, we discuss the investigated model Hamiltonian, the applied shuttling protocol, and the concept to measure leakage of the computational subspace of $\mathbb{Z}_4$ parafermions.
Our results are presented in Sect.~\ref{sect:results}, while their discussion is given in Sect.~\ref{sect:conclusion}.

\section{Methods}
\label{sect:methods}
We carried out  DMRG and  TDVP calculations using the two-site implementations of the ITensor program package~\cite{itensor, TDVP_ITensor}. 
Simulating a model with superconductivity and spin-nonconserving interaction, the only conserved quantum number to be considered is particle parity. Due to the sizable gap tuned by the parameters $\Delta$ and $V$, the DMRG calculations of instantaneous states are manageable for systems with hundreds of sites, as we demonstrated in Ref.~\onlinecite{Osvath_2024}.

Note that to reduce correlations in the system without altering the essential physics, we set open boundary conditions, which are favored by the DMRG approach. To minimize the distance between correlated sites in the DMRG chain mapping, the sites of the two legs follow each other in alternating order.
In our DMRG calculations, the bond dimension of the initial random matrices and the minimum bond dimension were set to $100$ while sweeping is allowed until a strict convergence in energy ($10^{-8}~t$) is reached. 
The excited states were obtained in an iterative fashion, where a projector of the lower-lying roots is added to the Hamiltonian with a penalty factor of magnitude $10~t$.
In TDVP simulations, time discretization is kept fixed at a value of $d\tau=0.1$, thus the number of iteration steps increases linearly with the shuttling time $T$. 

In Sect.~\ref{sect:modeling}, we also tested the effect of the truncation error $\varepsilon$,  which is the main source of numerical error of the applied computational methods, in great detail. Note that apart from the numerical parameters mentioned above, all other parameters of the ITensor codes were kept at their default value.
In the following, we briefly summarize the background of the applied methods, focusing on the discussion of the most critical numerical errors.

\subsection{DMRG}
We study the instantaneous ground state manifold of the interacting ladder model using the DMRG approach~\cite{White-1992b,White-1993}, which provides an approximate solution with controllable accuracy as described below. For more details, we recommend reading Ref.~\onlinecite{Schollwock-2011}, a seminal paper in the field.

The method iteratively optimizes the wave function represented in the so-called matrix product state (MPS) form.
In the MPS, a given matrix is assigned to each occupation state of a site, and the weight of a given configuration of the wave function can be obtained by multiplying these matrices. 
In each DMRG step, only matrices of two adjacent sites are updated simultaneously to minimize the energy while keeping the matrices of all other sites fixed. 
Most importantly, by factorizing the updated tensors using the singular value decomposition (SVD) approach, the dimension of the MPS matrices can be controlled. 

Recall that the singular values of the resulting Schmidt decomposed state, $\lambda$, satisfy the relation $\sum_{i} \lambda_i^2 = 1$. 
In practice, by keeping only the $m$ most dominant singular vectors (with the largest singular values) to act as the basis of the MPS matrices, their dimension can be efficiently truncated to a computationally manageable level with minimal loss of accuracy.
The discarded weights introduce the truncation error, $\varepsilon = 1 - \sum_{i\leq m} \lambda_i^2 \ll 1$.
In our calculations, in each DMRG step, the dimension of the truncated matrix (so-called bond or link dimension) is adjusted to keep the $\varepsilon$ truncation error fixed~\cite{Legeza-2003a}.

\begin{figure*}[t]
    \centering
    \includegraphics[width=\linewidth]{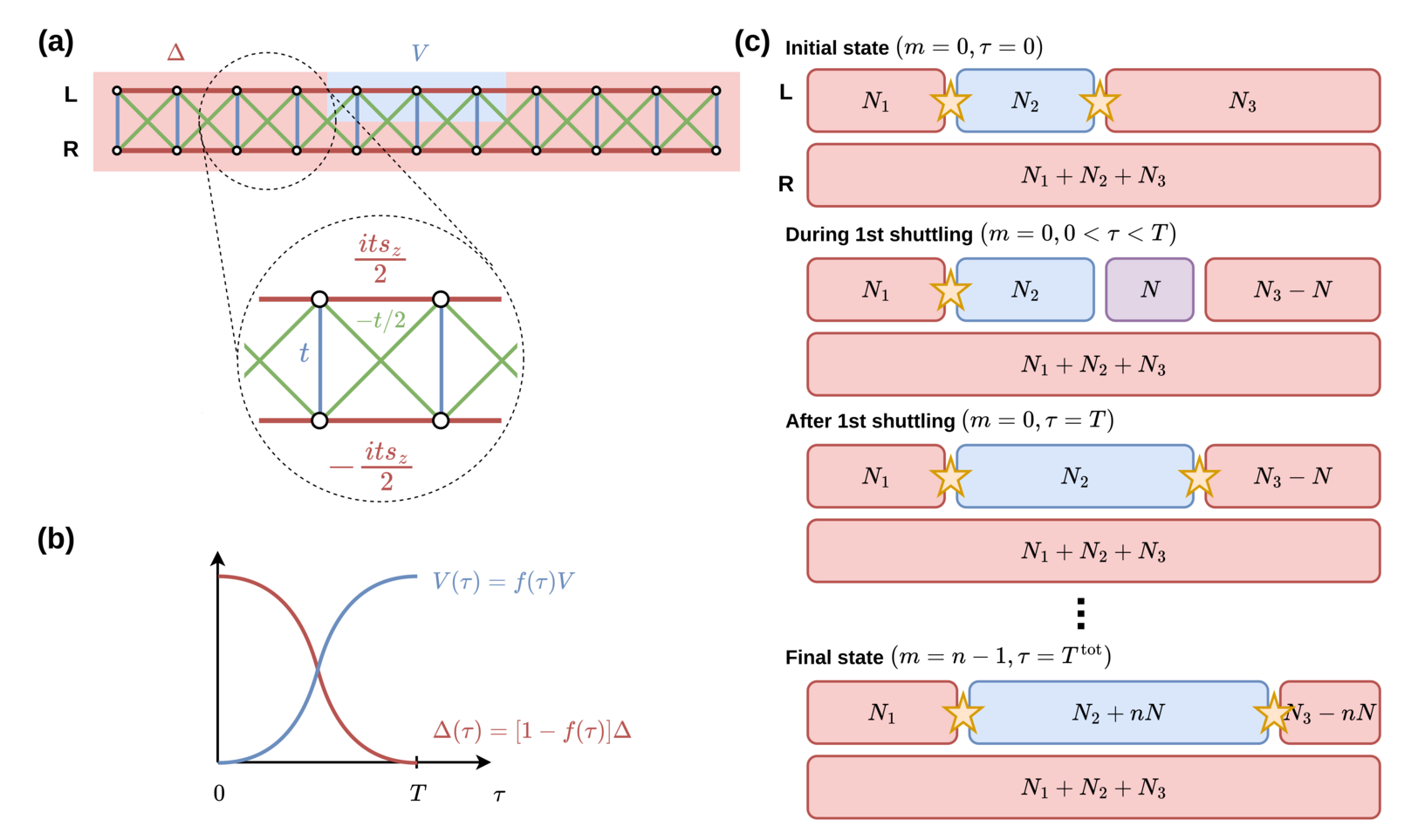}
    \caption{%
    Visualization of the model and the shuttling protocol. 
    (a) Sketch of the kinetic term and the investigated model configuration setup. 
    In the figure,  the black circles denote the ladder sites. Light blue and red shades denote the region with superconductivity with a quasiparticle gap $\Delta$ and with interaction of strength $V$, respectively. 
    The dark blue, red and green lines connecting the black circles represent the distinct hopping processes, which are detailed in the zoomed picture.
    (b) Depiction of the smooth evolution between two configuration in a single piano-key stroke (over time $T$) is tuned as defined in Eq.~\ref{eq:shuttle_protocol}. 
    (c) Sketch of the transport carried out by the piano-key protocol. 
    In the schematic figures, the ladder sites are not explicitly visualized, but the length of the various ladder segments is denoted. 
    Starting from a model setup $(N_1, N_2,N_3)$, where $N_2$ is the length of the interacting segment while $N_1$ and $N_3$ are the superconducting terminal lengths, in each shuttling step, we move one of the domain walls $N$ sites further into the superconducting region over time $T$.   
    The yellow star denotes the position fof the interface-localized parafermion state. Note that the parafermionic state is not localized in the $N$ block during shuttling hence we omit that star. 
 }
    \label{fig:shuttle_protocol}
\end{figure*}

\subsection{TDVP}
Time-evolving a strongly correlated quantum system can be a challenging task due to the drastic growth of entanglement over time.
To tackle this, we employ the TDVP approach~\cite{Haegaman_TDVP_PhysRevLett.107.070601,Haegeman2013,Haegeman2016}, which allows us to efficiently simulate the time-evolution of an initial state expanded in MPS form.
In particular, the TDVP method projects the time-evolution operator onto the tangent space of the MPS manifold, ensuring that the evolved state remains within the bounds of the specified MPS ansatz.

The evolution from  $\tau=0$ to $\tau=T$ is performed in discrete time steps of size $d \tau$, where each step is evaluated using a DMRG-like sweep.
Therefore, the TDVP calculation is performed in total $N_\text{sw} = \frac{T}{d \tau}$  sweeps.
The TDVP sweep essentially implements the second-order Trotter-Suzuki decomposition~\cite{Hatano2005} of the time-evolution operator. 
In each step of the sweep, for a given pair of neighboring sites in the MPS chain, we solve the projected local Schrödinger equation for the $d\tau/2$ increment, applying the Krylov exponentiation~\cite{Saad1992}.
In the two-site variant of the TDVP approach, the evolved state given in MPS tensor format is truncated by SVD as in standard two-site DMRG, which is followed by backward propagation on the center site to prevent overcounting evolution.

Implementing the second-order Suzuki-Trotter decomposition in TDVP, the total integration error scales as $d\tau^{2}$ while the error induced by the projection can be controlled by the MPS bond dimension. 
Note that the two-site variant of TDVP allows for dynamic adjustment of the bond dimension, making it more suitable to capture entanglement growth during time evolution, albeit at an increased computational cost compared to the cheaper one-site alternative.
As in the DMRG method, discussed above, the accuracy of the TDVP approach is essentially controlled by the cutoff parameter $\varepsilon$. 

In practice, the step-halving convergence analysis implied that the standard setting $d\tau=0.1$ provides sufficient precision in the studied diabatic processes.
To assess the actual effect of the truncation in our simulations, we performed  DMRG and TDVP calculations for various truncation errors in the range of $10^{-8}-10^{-5}$ as discussed in Sect.~\label{sect:results}.

\section{Measuring leakage during shuttling}
\label{sect:modeling}
In this section,  we present the investigated model Hamiltonian and the applied shuttling protocol. We also discuss a potential generalization of the diabatic error induced by finite-time development for degenerate states, called diabatic leakage~\cite{leakage_PhysRevA.97.032306}.

\subsection{Model Hamiltonian and transport protocol}
The considered effective model of a two-leg ladder of spinful electrons aims to capture the physics of $\mathbb{Z}_4$ parafermion modes emerging at the interface of interacting and superconducting regions.
 
The kinetic term describes nearest-neighbor hopping of electrons given by
\begin{equation}
\begin{aligned}
\label{eq:h_kin}
    H_{\text{kin}} 
        =&\sum_{m } c_{m}^{\dagger} \left( -\mu_m s_0 \otimes \zeta_0 + t s_0 \otimes \zeta_x \right)c_{m} \\
        &-\frac{t}{2}\sum_{m} c_{m+1}^{\dagger} \left( i s_z \otimes \zeta_z +  s_0 \otimes \zeta_x \right)c_{m} + \hc \,.
\end{aligned}
\end{equation}
Here, $s,s'$ are spin indices, $\zeta \in \{\Lleg, \Rleg\}$ leg indices, and $c_{m,\zeta,s}$ creates an electron of spin $s$ on site $m$ of leg $\zeta$. The parameter $t$ sets the energy scale and $\mu_{m,\zeta}$ denotes the chemical potential. 
As noted earlier, in the low-energy limit, propagation along the two legs is decoupled by construction; thus, the model mimics the propagation of low-energy helical particles of spatially separated edges in a two-dimensional topological insulator.~\cite{Osvath_2024}.
An on-site $s$-wave pairing, 
\begin{equation}
\label{eq:h_sc}
    H_{\text{sc}}=\sum_{m \zeta}\Delta_{m,\zeta}\!\left(c_{m,\zeta,\up}^{\dagger}c_{m,\zeta,\down}^{\dagger}+\hc\right)\,,
\end{equation}
with amplitude $\Delta_{m,\zeta}\neq0$ is allowed only in the superconducting region where explicit electron-electron interactions are not considered.
In the interacting region, in contrast, $\Delta_{m,\zeta}=0$ is set and we apply a simple short-range two-electron interaction,  
\begin{equation}
\label{eq:h_int}
    H_{\text{int}}=\sum_{m ,\zeta} V_{m,\zeta}\!\left(c_{m,\zeta,\up}^{\dagger}c_{m,\zeta,\down}c_{m+1,\zeta,\up}^{\dagger}c_{m+1,\zeta,\down}+\hc\right)\,,
\end{equation}
which describes an anisotropic symmetric exchange of strength $V_{m,\zeta}$ between neighboring sites on the same leg.
Combining these terms, the total Hamiltonian is given by
\begin{equation}
\label{eq:hamiltonian}
    H = H_{\text{kin}} + H_{\text{sc}} + H_{\text{int}}\,,
\end{equation}
where the actual system configuration of the interacting and superconducting regions parametrized by site-independent $V, \mu$ and $\Delta$ values is indicated by blue and red, respectively, in Fig.~\ref{fig:shuttle_protocol} (a).
In particular, the right leg of the ladder features superconductivity throughout its length, and the  central interacting segment of $N_2$ sites in the left leg is bounded by superconducting leads of $N_1$ and $N_3$ sites.
Recall that this configuration, denoted as $H^{(N_1,N_2,N_3)}$ in the following, is found to accommodate $\mathbb{Z}_4$ parafermion modes for a wide range of model parameters~\cite{Osvath_2024}.
Note that $\mu_m=0$ is set in our calculations.

\subsection{Finite-time parafermion transport protocol}
To shuttle diabatically parafermion edge states by  $N^{\rm tot}$ sites  using the  piano-key approach~\cite{mzm_shuttle_truong_1}, the process can be broken into multiple shuttling steps  as illustrated in Fig.~\ref{fig:shuttle_protocol} (c). In each shuttling operation, we smoothly shift the upper interface on the left leg  towards the superconducting terminal by tuning the model parameters continuously over time.
Specifically, complete transport over the $N^\text{tot}$ sites is performed in $n$ equidistant shuttling steps of $N$ sites $\left(N^\text{tot}=nN\right)$.
The $n$ shuttling steps, each performed in $T$ time, are executed consecutively.  Therefore, the overall shuttle procedure, which is initiated from $(N_1,N_2,N_3)$ setup to $(N_1,N_2+nN,N_3-nN)$ configuration, takes $T^\text{tot}=nT$ time.
Formally, the shuttling Hamiltonian of step $m\in\{0,\dots,n-1\}$, performed in the time frame of  $\tau \in (mT,(m+1)T)$, 
reads
\begin{equation}
\begin{aligned}
\label{eq:shuttle_protocol}
    H_{T,N}^m(\tau) = 
       &\left(1 - f_{T}^{m}(\tau)\right) H^{(N_1,N_2+mN,N_3-mN)}\\
    &+f^{m}_{T}(\tau)H^{(N_1,N_2+(m+1)N,N_3-(m+1)N)}\,,
    \end{aligned}
\end{equation}
where $f_{T}^{m}$ is an increasing function that meets the boundary conditions $f_{T}^{m}(mT)=0$ and $f_{T}^{m}((m+1)T)=1$.
In our simulations, we choose the smooth tuning function $f_{T}^{m}(\tau) = \sin^2\left(\frac{\pi (\tau-mT)}{2T}\right)$, illustrated in Fig.~\ref{fig:shuttle_protocol} (b),  which is expected to yield a favorable small diabatic error, as demonstrated for transporting Majorana fermions~\cite{mzm_shuttle_truong_1}. 

Note that single-site shuttling ($N^\text{tot}=1$), discussed in the following, can be executed only in $n=1$ shuttling step ($T^\text{tot}=T$).
For compactness, in the following, the instantaneous ground state of $H_T(\tau)$ evaluated at $\tau=0$ and $T$ is indicated as $\Psi_0$ and $\Psi_T$, respectively.

\subsection{Quantification of diabatic leakage}
\label{sect:leakage}
The diabatic error of a time-evolved quantum system with a non-degenerate ground state quantifies the fidelity loss between its instantaneous final state $|\psi_T \rangle$ and the state propagated from the initial state $|\psi_0 \rangle$ over time $T$ by the operator $U(T)$ as 
\begin{equation}
    r(T) = 1 - |\langle \psi_T | U(T)| \psi_0 \rangle | ^2\,.
    \label{eq:diab_err_non_deg}
\end{equation}
It benchmarks the overall accuracy of the diabatic process, which is governed by the rate of the transition and the energy gap of the system.
This measure is typically applied for MZMs~\cite{Bauer2018}, where there is no topological degeneracy within the parity sectors.

In contrast, in the considered model, 2-2 odd and even degenerate $\mathbb{Z}_4$ egde states define the computational manifold. 
Consequently, an extension of the formula  Eq.~\ref{eq:diab_err_non_deg} given for pure states is needed to describe the leakage of the computational subspace into the bulk states during a diabatic process.

Assume a $n$-degenerate manifold characterized by instantaneous states  $\bm{\psi}_T=\{\psi_T^i\}_{i=1}^{n}$ and $\bm{\psi}_0=\{\psi_0^i\}_{i=1}^{n}$ at time $T$ and $0$, respectively.
The leakage projector,
\begin{equation}
   P_T =  1-\sum_i|\psi_{T}^i \rangle\ \!\!\! \langle \psi_T^i | \,,
   \label{eq:projector}
\end{equation}
which is orthogonal to the computational subspace spanned by the final states obtained at time $T$, forms the pillar of the analysis~\cite{leakage_PhysRevA.97.032306}.
The time-evolution of the maximally mixed initial states  $\left(\rho_0=1/n\sum_i|\psi_{0}^i \rangle\ \!\!\! \langle \psi_0^i |\right)$, propagated  over time $T$ by the operator $U(T)$, might induce a leakage that is measured using the projector of Eq.~\ref{eq:projector} as 
\begin{equation}
   L(T) =  \operatorname{Tr}( P_TU(T)\rho_0U(T)^\dagger)  = 1 - \frac{1}{n}\sum_{ij} |\langle \psi_T^i | U(T)| \psi_0^j \rangle | ^2.
\end{equation}
Here, $L$ ranges from 0 (no error) to 1 (maximal error) and returns the formula  Eq.~\ref{eq:diab_err_non_deg}  when $n=1$.
Note that leakage error is critical to be controlled by the speed of the shuttling processes in order to avoid a fatal loss of quantum information that cannot be retrieved by standard error correction.
\begin{figure*}[t]
    \centering
    \includegraphics[width=0.95\linewidth]{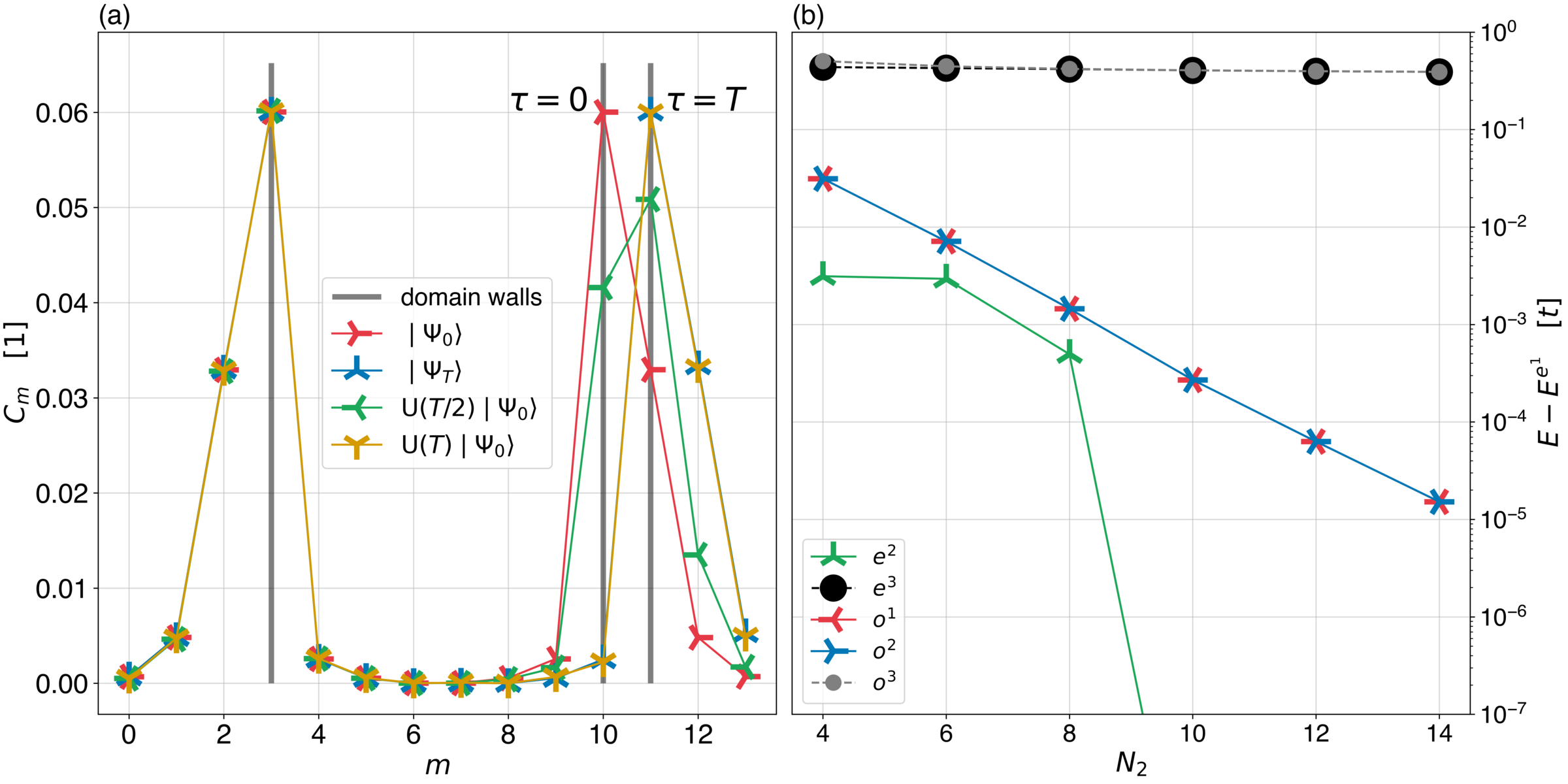}
    \caption{%
         Analysis of the parafermion bound states. a) 
         Transition strength $C_m$ for site $m$ of the left leg is defined in Eq.~\ref{eq:C}. Results, computed for initial system setup  $(N_1,N_2,N_3)=(3,8,3)$ at $V/t=2.2$, are presented for instantaneous states at initial ($\tau=0~\frac{1}{t}$) and final moment ($\tau=T=50~\frac{1}{t}$) of the single-site shuttling and for time-evolved states  $U(T/2)|\bm{\Psi}_0\rangle$, $U(T)|\bm{\Psi}_0\rangle$.
         b) Excitation spectrum  as a function of interaction subsystem length $N_2$. States denoted as $e^1$, $e^2$, $o^1$ and $o^2$ form $\bm{\Psi}_0$ the bound states with even and odd parity, while the $e_3$ and $o_3$ states are the lowest-lying bulk excitations.  The  DMRG data, presented for $\varepsilon=10^{-9}$ cutoff and $10^{-8}~t$ energy convergence condition, was obtained keeping the superconducting terminals of fixed length ($N_1=N_3=3$). 
    }
\label{fig:parafermion_check}
\end{figure*}
\begin{figure*}[t]
    \centering
    \includegraphics[width=0.95\linewidth]{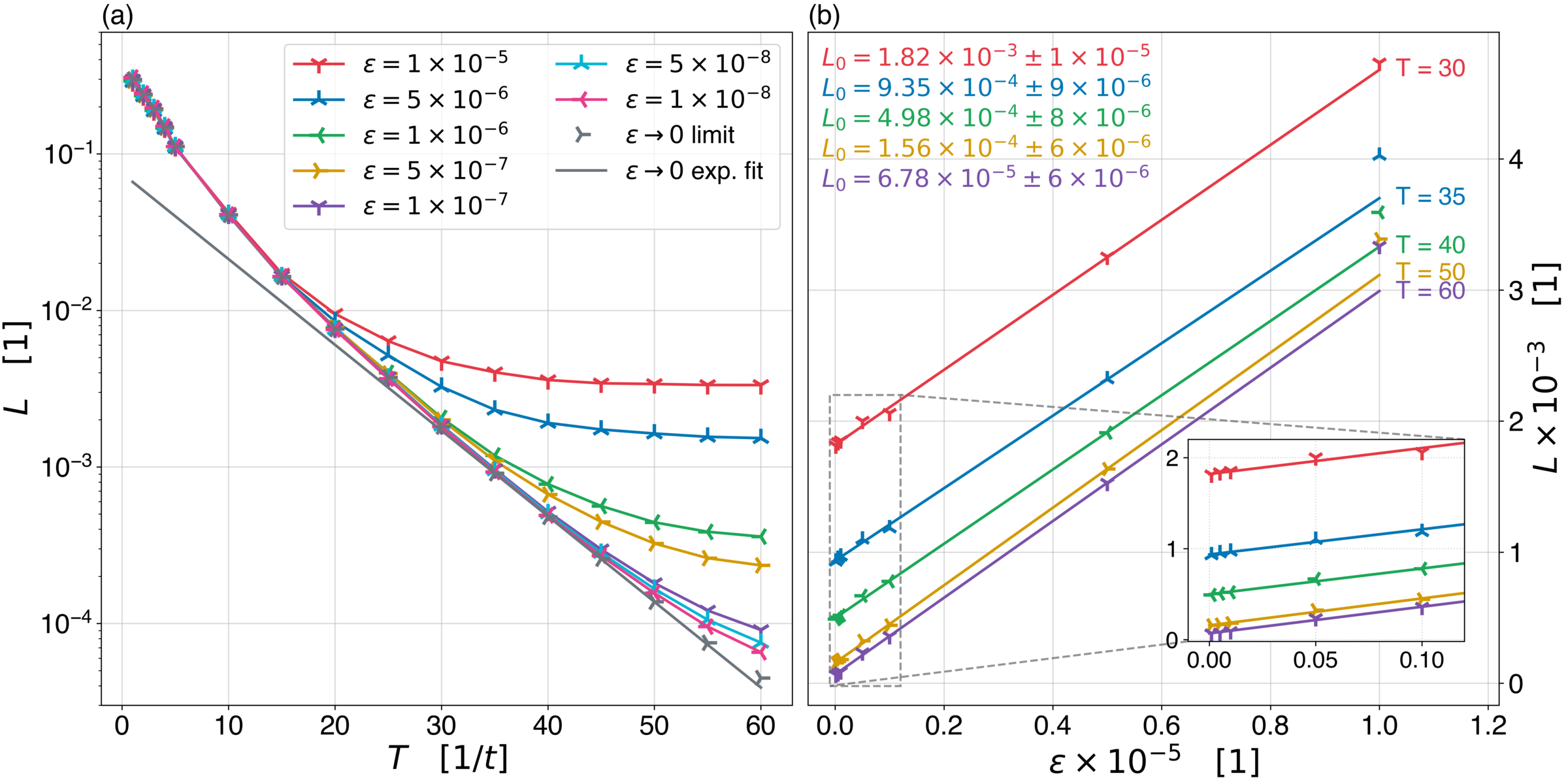}
    \caption{%
        Leakage induced by a diabatic single-site shuttling. (a) Numerical results of leakage $L$, obtained at various values of cutoff parameter $\varepsilon$, as a function of shuttling time $T$  for configuration $(N_1,N_2,N_3)=(3,8,3)$ at $V/t=2.2$ and $\Delta=1$. Colored curves are guide to the eye. The data for $\varepsilon \to 0$ limit is obtained from a linear extrapolation as discussed in the main text and visualized in panel b). Solid gray line is the exponential curve, fitted to the cutoff-free gray data, corresponding to the Landau-Zener theory. 
        (b) Diabatic leakage $L$ w.r.t. cutoff $\varepsilon$ for selected $T$ values. Raw data and fitted curves are shown up to $\varepsilon =10^{-5}$, while that particular data point of largest truncation error is omitted from the linear fitting. The infidelity extrapolated in the  $\varepsilon \to 0$ limit ($L_0$) is written in the color of the corresponding evolution time $T$.   
    }
\label{fig:T_vs_err_log}
\end{figure*}

\section{Results}
\label{sect:results}
In this section, we examine the impact of numerical errors of the applied computational approaches on  diabatic leakage for a single-site shuttling. 
We also discuss information loss in the case of different strategies for shuttling over multiple sites, and we also demonstrate the robustness of the edge states against leakage error across a range of model parameters.

\subsection{Differentiation of simulation and diabatic errors}
To ensure the reliability of our theoretical results, the numerical control parameters must be carefully tuned, minimizing their influence on the predicted leakage $L$. 
Correspondingly, in the following, we investigate the impact of the numerical errors of simulating single-site shuttling that are discussed in Sect.~\ref{sect:methods}. 
We focus on a suitably small, yet physically representative system that can be studied with varying simulation precision, practically allowing one to approach  the limit free of computational error.
Consequently, we study the configuration $(N_1,N_2,N_3) =(3,8,3)$ for a set of model parameters that is representative of the paramagnetic phase ($\mu=0,~\Delta=1,~V=2.2$ measured in $t$).

First, we demonstrate that the chosen model setup is large enough to describe the four topological edge states by analyzing the structure of the simulated ground state manifold and the system-size dependence of the energy spectrum.
Specifically, we study the transition strength of the local annihilation operator on the left leg between the odd and even  parafermionic manifolds. For the maximally mixed density matrices of edge states with even $\left(\{|e^i\rangle\}_{i=1}^2\right)$ and odd parity $\left(\{|o^i\rangle\}_{i=1}^2\right)$, the transition strength at site $m$, 
\begin{equation}
C_m= \frac{1}{4}\sum_{ijs}\left|\langle o^i | c_{m,L, s} | e^j \rangle\right|^2\,,
\label{eq:C}
\end{equation}
is expected to be sharply localized at the two interfaces~\cite{Osvath_2024}.
In Fig.~\ref{fig:parafermion_check}, we present $C_m$ for instantaneous eigenstates computed before and after shuttling ($\tau=0$ and $\tau=T$), denoted compactly by $|\bm{\Psi}_0\rangle$  and  $|\bm{\Psi}_T\rangle$, respectively.
As an illustration of the diabatic procedure performed in $T=50~\frac{1}{t}$ time, we also depict the propagated states at $\tau=T/2$  and $\tau=T$ , i.e.,  $U(T/2)|\bm{\Psi}_0\rangle$ and $U(T)|\bm{\Psi}_0\rangle$, respectively.
It is conspicuous that all simulated instantaneous states are localized at the interfaces of the corresponding configuration.
In fact, the matrix element is maximal at the interfaces, reaching a value of around 0.25, while it decays rapidly with distance, and in the center of the interacting region it is orders of magnitude smaller ($C_6\approx10^{-4}$).
Turning the focus to the time-evolved states, we find that the states evolved until half of the entire transport step ($\tau=T/2$) shows a smeared overlap distribution $C_m$ that is a combination of the initial and final instantaneous states.
Most importantly, at the end of the shuttling process ($\tau=T$), the profile of the transition strength of the evolved states $U(T)|\bm{\Psi}_0\rangle$ completely overlaps with the result for the instantaneous states of the final shuttled Hamiltonian $\left(|\bm{\Psi}_T\rangle\right)$.
 
In order to give a more qualitative description of the identified edge states, we study their energetics as a function of system size in more detail. 
In Fig.~\ref{fig:parafermion_check} (b), we show the lowest-lying energies relative to the ground state with even parity (dubbed $e^1$) for varying the length of the interaction segment, $N_2=4-14$, but keeping the terminal lengths fixed.
Most importantly, we find that the bulk gap (energy of $e^3$ and $o^3$ states in the figure) is as large as 0.4~$t$ irrespective of the distance between the two  interfaces. The invariance of the gap implies that the bulk excitations are the result of local disturbances of the ground state~\cite{fendley_parafermion}.
The energy of the two parafermion edge states with odd parity ($o^1$ and $o^2$), which form a Kramers pair, equal up to the precision of the modeling ($10^{-7}~t$). Their energy relative to the even ground state decays exponentially with increasing system size in line with theory~\cite{Teixeira_2022}. 
In contrast, the energy split between the bound states with even parity ($e^1$ vs $e^2$) induced by hybridization~\cite{Teixeira_2022}, is noticeable for small system sizes ($N_2=4-8$); nevertheless, it is significantly smaller than the odd-even splitting.

Altogether, the maximum energy split of the four bound states, which can potentially determine the dynamic phase induced in the shuttling processes, decays rapidly with system size, reaching $10^{-3}~t$ and $10^{-4}~t$ for $N_2=8$ and 12, respectively.
Based on these trends, we conclude that a detailed analysis of the setup of $N_1=N_3=3$ and $N_2=8$ might provide a reasonable compromise between the computational demands of the TDVP time evolution and the tangible representation of the $\mathbb{Z}_4$ edge states.

As part of the convergence analysis of the TDVP transport simulations, we investigated the impact of the preset truncation error $\varepsilon$ on diabatic leakage while keeping the moderate $d\tau=0.1$ setting fixed (see Fig.~\ref{fig:T_vs_err_log} (a)).
Physical intuition dictates that $L(T)$ decreases monotonically and rather rapidly  as a function of the drive time $T$ approaching the adiabatic limit. However, in practice, we find that the TDVP results obtained at finite truncations, e.g.,  $\varepsilon=10^{-5}$, do not fully meet anticipation, which can be understood as follows.
For sufficiently small values of $T$, the corresponding $L(T)$ is so large ($\sim 10^{-1}$) that it essentially hides the errors of the simulation and seems to be insensitive to the choice of truncation. 
In contrast, for higher values of $T$, where leakage decreases by orders of magnitude, the precision of the simulations becomes more critical, as the $L(T)$ calculated for a given $\varepsilon$ might be dominated by the truncation error and the curve starts to saturate.
In order to eliminate the truncation error, we fit the numerical infidelity data for a given $T$  as a function of $\varepsilon$ (see Fig.~\ref{fig:T_vs_err_log} (b)). As the defined infidelity quantities practically measure the discarded weight of the simulated wave functions along the leakage, linear fitting of the  data obtained for finite cutoffs is reasonable, just as in the case of typical DMRG energy extrapolation~\cite{Legeza1996}.
In Fig.~\ref{fig:T_vs_err_log} (a), the actual diabatic leakage is also shown, i.e., the result extrapolated to the truncation-free limit ($\varepsilon \rightarrow 0$). We find that the extrapolated leakage decreases monotonically with $T$ and starting from around $T=30~\frac{1}{t}$ shows exponential decay predicted by the Landau-Zener theory. 

\subsection{Diabatic leakage induced by shuttling}
\begin{figure}[h]
    \centering
    \includegraphics[width=0.95\linewidth]{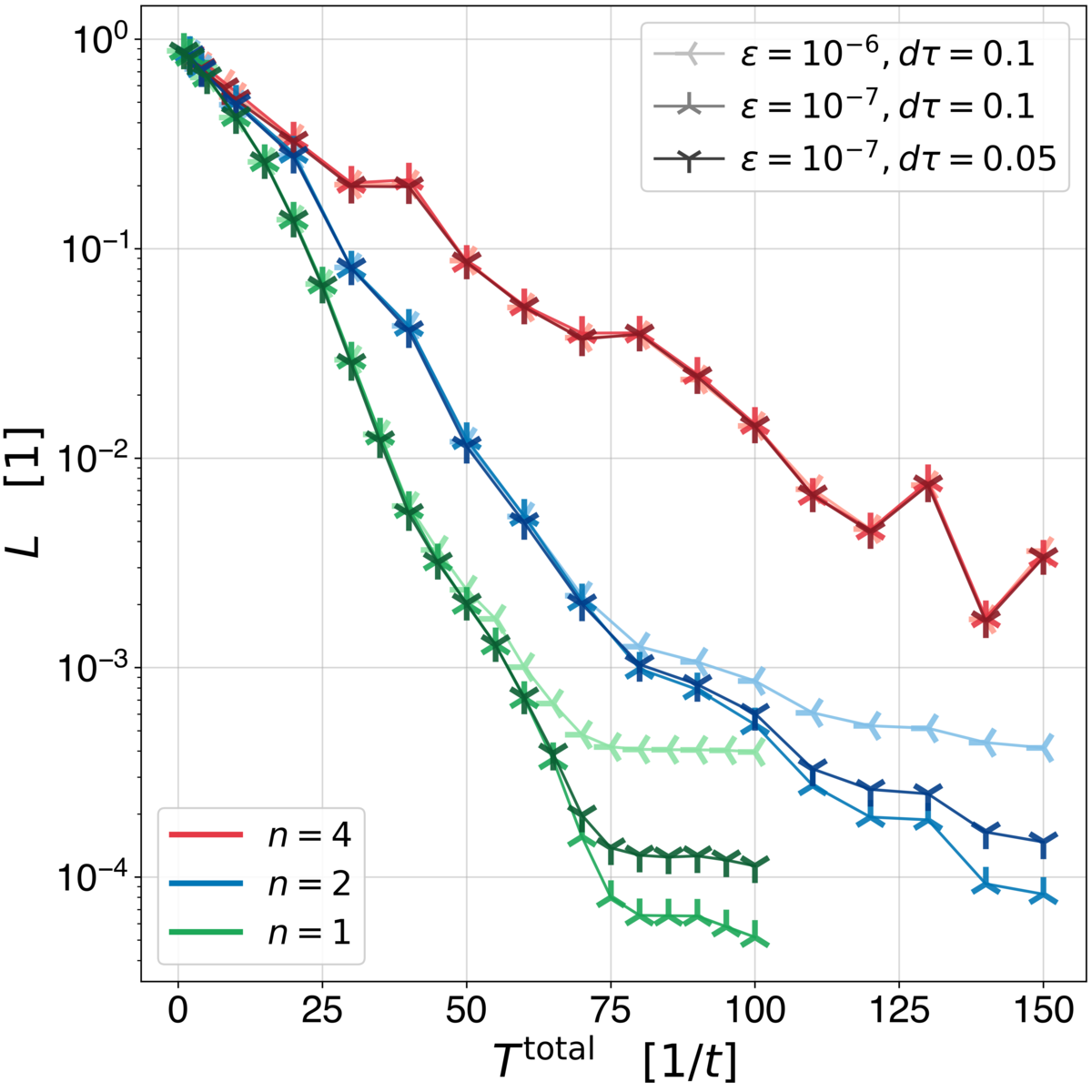} 
    \caption{%
    Leakage in case of diabatic four-site transport. Leakage $L$ as a function of total evolution time $T^\text{total}$ was obtained  using $n=1,2,4$ equally-sized piano-key shuttling protocol for initial configuration $(N_1,N_2,N_3)=(3,8,6)$ at $V/t=2.2$ and $\Delta/t=1$. Numerical control parameters ($\varepsilon$, $d\tau$) were varied to test the reliability of the simulations. The labeling is given in split-legend, where the distinct colors represent the different piano-key setups while the  hue of the color and the symbol encode the applied parametrization of the TDVP calculations.
    }
\label{fig:evolve_multiple_steps}
\end{figure}
\begin{figure*}[!bt]
    \centering
    \includegraphics[width=\linewidth]{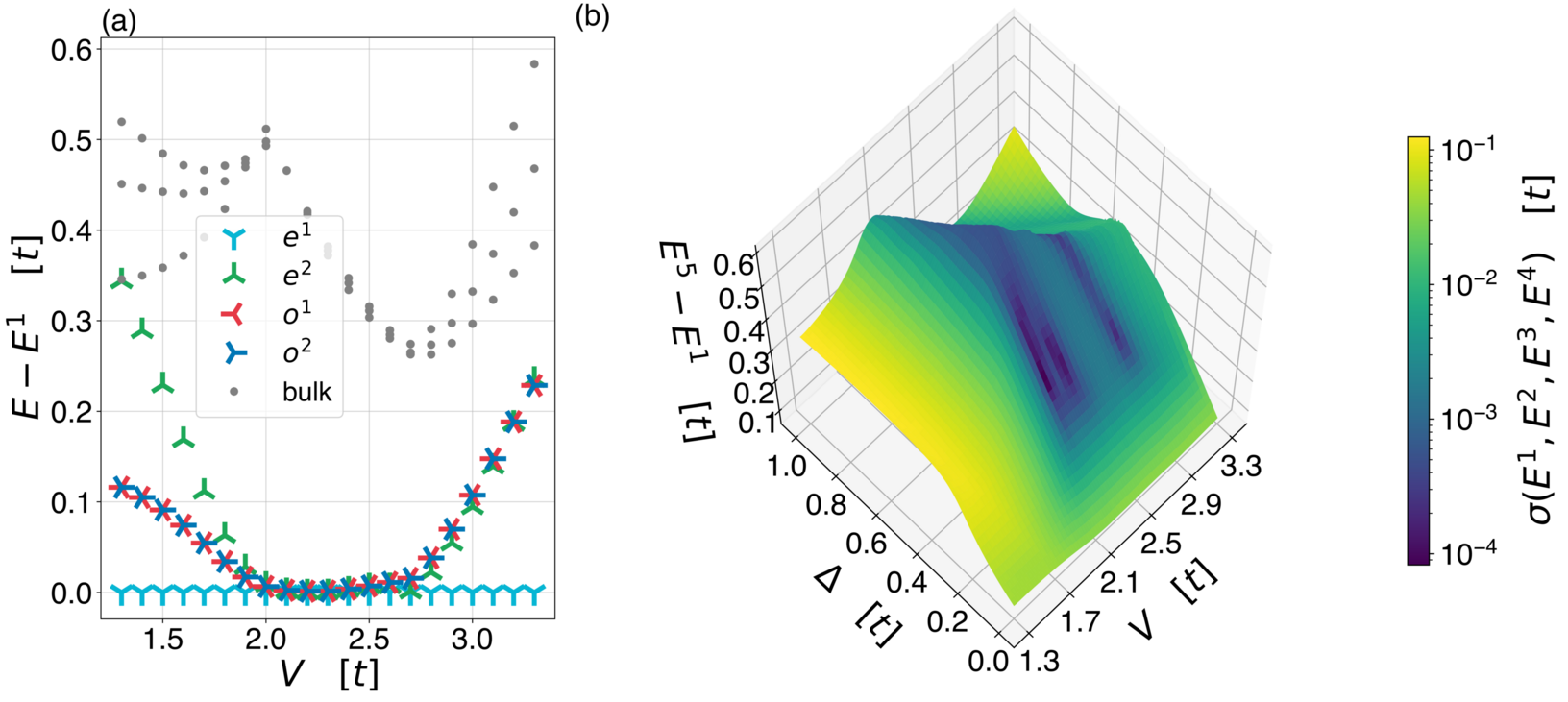}
    \caption{%
        Stability of the parafermion phase. (a) Excitation spectrum  as a function of interaction strength $V$ for the system setup (3,8,3), at fixed $\Delta/t=1$.
        (b) 4-dimensional plot of standard deviation of the 4 lowest lying states' energies and 5th state's excitation energy as a function of $\Delta$ and $V$ for the system setup (3,8,3). The $z$ coordinate represents the magnitude of the excitation energy that is the bulk-gap in the parafermionic region. The color encodes $\sigma(e^1,e^2,o^1,o^2)$, the closer it is to zero, the stronger the quasi-degeneracy is.
    }
\label{fig:gap_analysis}
\end{figure*}
Following the numerical analysis of the TDVP-based simulation of single-site shuttling, we turn the focus to the transport of the parafermion mode over multiple sites.
Specifically, we investigate the leakage of a four-site shuttling executed on the right terminal of the initial setup $(N_1,N_2,N_3)=(3,8,6)$ at model parameters $V/t=2.2$ and $\Delta/t=1$.
The diabatic leakage accumulated during transport over $N^\text{tot}=4$ sites was tested for $N=1,2,$ and 4-site piano-key shuttling protocols discussed in \ref{sect:leakage}.
The corresponding numerical results obtained for various total shuttling times are visualized in Fig.~\ref{fig:evolve_multiple_steps}.
In particular,   we find that the leakage becomes sensitive to the choice of the truncation error ($\varepsilon =10^{-6}-10^{-7}$)  around $L\approx10^{-3}$ in line with the  observations on single-site shuttling studied above.
More importantly, we find that, in this demonstrative analysis of four-site transport, single-key diabatic shuttling ($n=1$) results in the lowest leakage  with rather smooth decay in the investigated range of drive-time.
In addition to cutoff analysis, the simulations were also tested for sensitivity to the choice of time-discretization as shown in Fig.~\ref{fig:evolve_multiple_steps}. All these numerical convergence tests result practically in the same decaying pattern until around $L\approx10^{-3}$. Consequently, the fluctuation observed in the case of $n=2$ and especially $n=4$ is not an artifact of the simulations but an actual physical effect attributed to the Landau-Zener-Stückelberg interference. According to theory~\cite{Shevchenko_2010}, in a non-adiabatically evolving system  the different components of the diabatic wavefunction accumulate relative dynamical phases. Therefore, in the case of transport protocols consisting of a consecutive series of shuttling steps (increasing $n$), the buildup of interference could cause more pronounced oscillations in the leakage curves compared to the $n=1$ case.
Note that similar patterns can also be observed on larger scales; see, for example, Ref.~\citenum{mzm_shuttle_truong_1}.

Based on the above findings at $V/t = 2.2$ and $\Delta/t=1$, which reveal an exponential decrease in the leakage governed by the bulk gap, we also explored the potential scope of this rapid process by analyzing the energetics in an extensive range of model parameters.
In Fig.~\ref{fig:gap_analysis} (a), we present the lowest-lying relative energies tuning $V/t$ between 1.3 and 3.3 while keeping $\Delta/t=1$ fixed. The fourfold (quasi)degeneracy of the parafermion ground state  is demonstrated for  interaction strength in the range of around 2.0-2.7.
Furthermore, we also find that these edge states are separated from the bulk states by a substantial gap of 0.3-0.5~$t$ for the investigated parameters. 
We also extended the analysis for various values of $\Delta/t$ in the range of 0.0-1.0.
For the sake of clarity, in Fig.~\ref{fig:gap_analysis} (b), we analyze the large amount of data by the standard deviation $\sigma$ of the four lowest-lying energies, as a measure of their degeneracy, and by the excitation energy of the fifth state, which defines the bulk gap in the parafermionic phase.
The fourfold (quasi)degenerate parafermionic phase, characterized by a spread well below  $0.1~t$, can be detected in the range of 0.2-1.0 superconducting coupling strength and is also stabilized for higher values of $V$ compared to the case of  $\Delta/t=1$.
In the whole parafermionic phase of the investigated model, the bulk gap is sizeable and varies in the range of 0.2-0.6~$t$. 
These energetic results imply the potential feasibility of transporting parafermionic modes with reasonable speed over a wide range of model parameters.

Finally, establishing a characteristic time scale that preserves the parafermionic subspace  during the shuttling is critical to assess an upper bound of operational frequency in practice.
As a point of reference, an adiabatic noise below $10^{-4}$ might be considered as a conservative upper bound estimate for fault-tolerant operation~\cite{steane1999,knill2005}.
Based on our extrapolated results in Fig.~\ref{fig:T_vs_err_log} (a), we estimate that $T^*=51~\frac{1}{t}$ is already large enough for reasonably accurate single-site shuttling ($L^*\approx 1\times10^{-4}$)  for the given model setup.
Based on results for multi-site shuttling in Fig.~\ref{fig:evolve_multiple_steps}, we might anticipate even lower but comparable estimate for critical shuttling time per site ($T^*$). 
Note that a very similar $T^*$ is expected for larger system sizes as well considering that the diabatic leakage follows the Landau-Zener theory governed by the bulk-gap, which is found to be practically size independent (see Fig.~\ref{fig:parafermion_check} (b)). 

In order to convert the critical $T^*$ to an actual physical time scale, the value of the parameter $t$ in physical realizations must be first estimated. 
In recent decades, among other concepts, gate-based and atomic approaches have been intensively developed to implement tunable quantum dots on Si and GaAs wafers~\cite{Burkard_2023}. Depending on the actual realization, the covered range of the hopping amplitude varies greatly; nevertheless, a wide bound of 0.01-10 meV can currently be given in overall. It is notable that that technically more challenging higher end~\cite{Kiczynski_2022} corresponds already to the sub-picosecond timescale. 
Therefore, setting a reasonable value of $t=0.5$~meV, a negligible diabatic leakage ($L^*\approx10^{-4}$) might be reached for a single-site shuttle around a critical time of $T^*=67$~ps (see Fig.~\ref{fig:T_vs_err_log}). Comparable critical shuttling time per site can be estimated in  the case of multiple-site transport (see Fig.~\ref{fig:evolve_multiple_steps}).
Note that in the case of MZMs the critical $10^{-4}$ diabatic error of shuttling was also found to be reached within rather similar time-scale (around 10 ps per site on average) assuming a hopping amplitude as large as 1.5 meV and using optimized transport protocols~\cite{mzm_shuttle_truong_1}.
More importantly, from the point of view of quantum dot arrays, shuttling of a particle with reasonable fidelity can typically be executed in the order of magnitude of nanoseconds per site~\cite{Baart_2016,Mills2019,DeSmet2025}.
The experiments are essentially limited by the frequency of the controlling electronics, the scaling limitations of synchronization, and by the fabrication imperfections.
Since the estimated leakage time-scale of the considered model is shorter than typical control times in present quantum-dot shuttling experiments, 
the adiabatic requirement for the idealized model does not pose a bottleneck for current setups.

Finally, note that leakage during diabatic transport measures the irreversible loss of quantum information from the computational subspace, which determines the potential lower limit for the shuttling time while maintaining the critical adiabatic condition.
However, the potential duration of a fault-tolerant shuttling also has an upper limit constrained by quasi-particle poisoning and dephasing effects within the nearly degenerate manifold, which can cause decoherence.
The optimal frequency of shuttling operations is constrained by these simultaneous  limits.

\section{Conclusion}
\label{sect:conclusion}
In this exploratory numerical study, we investigated the resilience of finite-time shuttling $\mathbb{Z}_4$ parafermion modes in a quantum dot array to leakage from the computational manifold into bulk states.

For studying these strongly correlated quantum systems, we applied the DMRG method, which also allowed us to simulate  dynamic processes efficiently by employing the TDVP approach.
The precision of the approximate simulations was scrutinized to provide an accurate prediction of the energies and the fidelity loss of the finite-time shuttling.

Specifically, we investigated the single- and multi-site shuttling of the edge modes applying the diabatic piano-key  protocol.
We found that the sub-nanosecond single-site shuttling time of the $\mathbb{Z}_4$ parafermion edge states compares favorably with the findings for Majorana zero modes which can be attributed to the observed large bulk gaps.
Furthermore, our theoretical prediction might also indicate that the shuttling speed of parafermions in quantum dot arrays might be limited by the timescale of the actual physical realization of the shuttling protocol rather than by the model system itself.

Our theoretical findings, combined with practical engineering considerations, could advance parafermion-based braiding architectures in quantum dot arrays.

\section*{Acknowledgments}
O. L. was supported by the Ministry of Culture and Innovation and the National Research, Development and Innovation Office within the Quantum Information National Laboratory of Hungary (Grant No. 2022-2.1.1-NL-2022-00004), National Research, Development and Innovation Office (NKFIH) through Grant Nos. K134437 as well
as projects KKP133827 and K142179. 
This project is supported by the TRILMAX Horizon Europe consortium (Grant No. 101159646).
We acknowledge Digital Government Development and Project Management Ltd. for awarding us access to the Komondor HPC facility based in Hungary.

\bibliography{references} 

@article{TDVP_ITensor,
  title = {Time-dependent variational principle with ancillary Krylov subspace},
  author = {Yang, Mingru and White, Steven R.},
  journal = {Phys. Rev. B},
  volume = {102},
  issue = {9},
  pages = {094315},
  numpages = {6},
  year = {2020},
  month = {Sep},
  publisher = {American Physical Society},
  doi = {10.1103/PhysRevB.102.094315},
  url = {https://link.aps.org/doi/10.1103/PhysRevB.102.094315}
}

@Article{knill2005,
author={Knill, E.},
title={Quantum computing with realistically noisy devices},
journal={Nature},
year={2005},
month={Mar},
day={01},
volume={434},
number={7029},
pages={39-44},
issn={1476-4687},
doi={10.1038/nature03350},
url={https://doi.org/10.1038/nature03350}
}

@article{Saad1992,
author = {Saad, Y.},
title = {Analysis of Some Krylov Subspace Approximations to the Matrix Exponential Operator},
journal = {SIAM Journal on Numerical Analysis},
volume = {29},
number = {1},
pages = {209-228},
year = {1992},
doi = {10.1137/0729014},
URL = { 
            https://doi.org/10.1137/0729014
},
eprint = { 
        https://doi.org/10.1137/0729014
        }
}

@Inbook{Hatano2005,
author="Hatano, Naomichi
and Suzuki, Masuo",
editor="Das, Arnab
and K. Chakrabarti, Bikas",
title="Finding Exponential Product Formulas of Higher Orders",
bookTitle="Quantum Annealing and Other Optimization Methods",
year="2005",
publisher="Springer Berlin Heidelberg",
address="Berlin, Heidelberg",
pages="37--68",
isbn="978-3-540-31515-5",
doi="10.1007/11526216_2",
url="https://doi.org/10.1007/11526216_2"
}

@article{Shevchenko_2010,
title = {Landau–Zener–Stückelberg interferometry},
journal = {Physics Reports},
volume = {492},
number = {1},
pages = {1-30},
year = {2010},
issn = {0370-1573},
doi = {https://doi.org/10.1016/j.physrep.2010.03.002},
url = {https://www.sciencedirect.com/science/article/pii/S0370157310000815},
author = {S.N. Shevchenko and S. Ashhab and Franco Nori}
}

@Article{Kiczynski_2022,
author={Kiczynski, M.
and Gorman, S. K.
and Geng, H.
and Donnelly, M. B.
and Chung, Y.
and He, Y.
and Keizer, J. G.
and Simmons, M. Y.},
title={Engineering topological states in atom-based semiconductor quantum dots},
journal={Nature},
year={2022},
month={Jun},
day={01},
volume={606},
number={7915},
pages={694-699},
issn={1476-4687},
doi={10.1038/s41586-022-04706-0},
url={https://doi.org/10.1038/s41586-022-04706-0}
}

@article{Burkard_2023,
  title = {Semiconductor spin qubits},
  author = {Burkard, Guido and Ladd, Thaddeus D. and Pan, Andrew and Nichol, John M. and Petta, Jason R.},
  journal = {Rev. Mod. Phys.},
  volume = {95},
  issue = {2},
  pages = {025003},
  numpages = {58},
  year = {2023},
  month = {Jun},
  publisher = {American Physical Society},
  doi = {10.1103/RevModPhys.95.025003},
  url = {https://link.aps.org/doi/10.1103/RevModPhys.95.025003}
}

@article{Teixeira_2022,
  title = {Overlap of parafermionic zero modes at a finite distance},
  author = {Teixeira, Raphael L. R. C. and Haller, Andreas and Singh, Roshni and Mathew, Amal and Idrisov, Edvin G. and Dias da Silva, Luis G. G. V. and Schmidt, Thomas L.},
  journal = {Phys. Rev. Res.},
  volume = {4},
  issue = {4},
  pages = {043094},
  numpages = {9},
  year = {2022},
  month = {Nov},
  publisher = {American Physical Society},
  doi = {10.1103/PhysRevResearch.4.043094},
  url = {https://link.aps.org/doi/10.1103/PhysRevResearch.4.043094}
}

@article{Legeza1996,
  title = {Accuracy of the density-matrix renormalization-group method},
  author = {Legeza, \"Ors and F\'ath, G\'abor},
  journal = {Phys. Rev. B},
  volume = {53},
  issue = {21},
  pages = {14349--14358},
  numpages = {0},
  year = {1996},
  month = {Jun},
  publisher = {American Physical Society},
  doi = {10.1103/PhysRevB.53.14349},
  url = {https://link.aps.org/doi/10.1103/PhysRevB.53.14349}
}

@article{Safwan_2025,
doi = {10.1088/1751-8121/aded51},
url = {https://doi.org/10.1088/1751-8121/aded51},
year = {2025},
month = {aug},
publisher = {IOP Publishing},
volume = {58},
number = {33},
pages = {335302},
author = {Safwan, Ali Hamed and Bomantara, Raditya Weda},
title = {Generating non-Clifford gate operations through exact mapping between Majorana fermions and $\mathbb{Z}_4$ parafermions},
journal = {Journal of Physics A: Mathematical and Theoretical},
}

@inbook{steane1999,
author = {Steane, A. M.},
publisher = {John Wiley \& Sons, Ltd},
isbn = {9783527603091},
title = {Space, Time, Parallelism and Noise Requirements for Reliable Quantum Computing},
booktitle = {Quantum Computing},
chapter = {8},
pages = {137-151},
doi = {https://doi.org/10.1002/3527603093.ch8},
url = {https://onlinelibrary.wiley.com/doi/abs/10.1002/3527603093.ch8},
eprint = {https://onlinelibrary.wiley.com/doi/pdf/10.1002/3527603093.ch8},
year = {1999}
}

@article{Haegeman2016,
  title = {Unifying time evolution and optimization with matrix product states},
  author = {Haegeman, Jutho and Lubich, Christian and Oseledets, Ivan and Vandereycken, Bart and Verstraete, Frank},
  journal = {Phys. Rev. B},
  volume = {94},
  issue = {16},
  pages = {165116},
  numpages = {10},
  year = {2016},
  month = {Oct},
  publisher = {American Physical Society},
  doi = {10.1103/PhysRevB.94.165116},
  url = {https://link.aps.org/doi/10.1103/PhysRevB.94.165116}
}

@article{Haegaman_TDVP_PhysRevLett.107.070601,
  title = {Time-Dependent Variational Principle for Quantum Lattices},
  author = {Haegeman, Jutho and Cirac, J. Ignacio and Osborne, Tobias J. and Pi\ifmmode \check{z}\else \v{z}\fi{}orn, Iztok and Verschelde, Henri and Verstraete, Frank},
  journal = {Phys. Rev. Lett.},
  volume = {107},
  issue = {7},
  pages = {070601},
  numpages = {5},
  year = {2011},
  month = {Aug},
  publisher = {American Physical Society},
  doi = {10.1103/PhysRevLett.107.070601},
  url = {https://link.aps.org/doi/10.1103/PhysRevLett.107.070601}
}

@article{Haegeman2013,
  title = {Post-matrix product state methods: To tangent space and beyond},
  author = {Haegeman, Jutho and Osborne, Tobias J. and Verstraete, Frank},
  journal = {Phys. Rev. B},
  volume = {88},
  issue = {7},
  pages = {075133},
  numpages = {35},
  year = {2013},
  month = {Aug},
  publisher = {American Physical Society},
  doi = {10.1103/PhysRevB.88.075133},
  url = {https://link.aps.org/doi/10.1103/PhysRevB.88.075133}
}

@Article{Bauer2018,
	title={{Dynamics of Majorana-based qubits operated with an array of tunable gates}},
	author={Bela Bauer and Torsten Karzig and Ryan V. Mishmash and Andrey E. Antipov and Jason Alicea},
	journal={SciPost Phys.},
	volume={5},
	pages={004},
	year={2018},
	publisher={SciPost},
	doi={10.21468/SciPostPhys.5.1.004},
	url={https://scipost.org/10.21468/SciPostPhys.5.1.004},
}

@misc{itensor,
title={The \mbox{ITensor} Software Library for Tensor Network Calculations},
author={Matthew Fishman and Steven R. White and E. Miles Stoudenmire},
year={2020},
eprint={2007.14822},
archivePrefix={arXiv}
}

@article{alicea_fendley,
author = {Alicea, Jason and Fendley, Paul},
title = {Topological Phases with Parafermions: Theory and Blueprints},
journal = {Annual Review of Condensed Matter Physics},
volume = {7},
number = {1},
pages = {119-139},
year = {2016},
doi = {10.1146/annurev-conmatphys-031115-011336}
}

@article{fendley_parafermion,
	doi = {10.1088/1742-5468/2012/11/p11020},
	url = {https://doi.org/10.1088/1742-5468/2012/11/p11020},
	year = 2012,
	month = {nov},
	publisher = {{IOP} Publishing},
	volume = {2012},
	number = {11},
	pages = {P11020},
	author = {Paul Fendley},
	title = {Parafermionic edge zero modes in ${Z}_{n}$-invariant spin chains},
	journal = {Journal of Statistical Mechanics: Theory and Experiment}
}

@article{Kitaev_2001,
	doi = {10.1070/1063-7869/44/10s/s29},
	url = {https://doi.org/10.1070/1063-7869/44/10s/s29},
	year = 2001,
	month = {oct},
	publisher = {Uspekhi Fizicheskikh Nauk ({UFN}) Journal},
	volume = {44},
	number = {10S},
	pages = {131--136},
	author = {A Yu Kitaev},
	title = {Unpaired Majorana fermions in quantum wires},
	journal = {Physics-Uspekhi}
}

@article{zhang,
  title = {Time-Reversal-Invariant ${Z}_{4}$ Fractional Josephson Effect},
  author = {Zhang, Fan and Kane, C. L.},
  journal = {Phys. Rev. Lett.},
  volume = {113},
  issue = {3},
  pages = {036401},
  numpages = {5},
  year = {2014},
  month = {Jul},
  publisher = {American Physical Society},
  doi = {10.1103/PhysRevLett.113.036401},
  url = {https://link.aps.org/doi/10.1103/PhysRevLett.113.036401}
}

@article{hutter2016quantum,
  title={Quantum computing with parafermions},
  author={Hutter, Adrian and Loss, Daniel},
  journal={Physical Review B},
  volume={93},
  number={12},
  pages={125105},
  year={2016},
  publisher={APS}
}

@article{orth2015non,
  title={Non-Abelian parafermions in time-reversal-invariant interacting helical systems},
  author={Orth, Christoph P and Tiwari, Rakesh P and Meng, Tobias and Schmidt, Thomas L},
  journal={Physical Review B},
  volume={91},
  number={8},
  pages={081406},
  year={2015},
  publisher={APS}
}

@article{para_braiding_solofo,
  title = {Parafermion braiding in fractional quantum Hall edge states with a finite chemical potential},
  author = {Groenendijk, Solofo and Calzona, Alessio and Tschirhart, Hugo and Idrisov, Edvin G. and Schmidt, Thomas L.},
  journal = {Phys. Rev. B},
  volume = {100},
  issue = {20},
  pages = {205424},
  numpages = {17},
  year = {2019},
  month = {Nov},
  publisher = {American Physical Society},
  doi = {10.1103/PhysRevB.100.205424},
  url = {https://link.aps.org/doi/10.1103/PhysRevB.100.205424}
}

@article{Klinovaja_PhysRevLett.112.246403,
  title = {Parafermions in an Interacting Nanowire Bundle},
  author = {Klinovaja, Jelena and Loss, Daniel},
  journal = {Phys. Rev. Lett.},
  volume = {112},
  issue = {24},
  pages = {246403},
  numpages = {5},
  year = {2014},
  month = {Jun},
  publisher = {American Physical Society},
  doi = {10.1103/PhysRevLett.112.246403},
  url = {https://link.aps.org/doi/10.1103/PhysRevLett.112.246403}
}

@article{White-1993,
  author = {White, Steven R.},
  title = {Density-matrix algorithms for quantum renormalization groups},
  journal = {Phys. Rev. B},
  volume = {48},
  issue = {14},
  pages = {10345--10356},
  numpages = {0},
  year = {1993},
  month = {Oct},
  publisher = {American Physical Society},
  doi = {10.1103/PhysRevB.48.10345},
  url = {http://link.aps.org/doi/10.1103/PhysRevB.48.10345},
}

@article{Legeza-2003a,
  author = {Legeza, {\"O}. and R\"oder, J. and Hess, B. A.},
  title = {Controlling the accuracy of the density-matrix renormalization-group method: The dynamical block state selection approach},
  journal = {Phys. Rev. B},
  volume = {67},
  issue = {12},
  pages = {125114},
  numpages = {10},
  year = {2003},
  month = {Mar},
  publisher = {American Physical Society},
  doi = {10.1103/PhysRevB.67.125114},
  url = {http://link.aps.org/doi/10.1103/PhysRevB.67.125114},
}

@article{Schollwock-2011,
title = "The density-matrix renormalization group in the age of matrix product states",
journal = "Ann. Phys.",
volume = "326",
number = "1",
pages = "96 - 192",
year = "2011",
note = "January 2011 Special Issue",
issn = "0003-4916",
doi = "https://doi.org/10.1016/j.aop.2010.09.012",
url = "http://www.sciencedirect.com/science/article/pii/S0003491610001752",
author = {Ulrich Schollw\"ock}
}

@article{White-1992b,
  author = {White, Steven R.},
  title = {Density matrix formulation for quantum renormalization groups},
  journal = {Phys. Rev. Lett.},
  volume = {69},
  issue = {19},
  pages = {2863--2866},
  numpages = {0},
  year = {1992},
  month = {Nov},
  publisher = {American Physical Society},
  doi = {10.1103/PhysRevLett.69.2863},
  url = {http://link.aps.org/doi/10.1103/PhysRevLett.69.2863}
}

@article{Sarma2015,
  doi = {10.1038/npjqi.2015.1},
  url = {https://doi.org/10.1038/npjqi.2015.1},
  year = {2015},
  month = oct,
  publisher = {Springer Science and Business Media {LLC}},
  volume = {1},
  number = {1},
  author = {Sankar Das Sarma and Michael Freedman and Chetan Nayak},
  title = {Majorana zero modes and topological quantum computation},
  journal = {npj Quantum Information}
}

@article{Fendley_2014,
doi = {10.1088/1751-8113/47/7/075001},
url = {https://doi.org/10.1088/1751-8113/47/7/075001},
year = {2014},
month = {jan},
publisher = {IOP Publishing},
volume = {47},
number = {7},
pages = {075001},
author = {Fendley, Paul},
title = {Free parafermions},
journal = {Journal of Physics A: Mathematical and Theoretical}
}

@article{lr2b-nmrk,
  title = {Transport of Majorana bound states in the presence of telegraph noise},
  author = {Sahu, Dibyajyoti and Gangadharaiah, Suhas},
  journal = {Phys. Rev. B},
  volume = {111},
  issue = {23},
  pages = {235306},
  numpages = {13},
  year = {2025},
  month = {Jun},
  publisher = {American Physical Society},
  doi = {10.1103/lr2b-nmrk},
  url = {https://link.aps.org/doi/10.1103/lr2b-nmrk}
}

@Article{Liu2025,
author={Liu, Hui
and Perea-Causin, Raul
and Bergholtz, Emil J.},
title={Parafermions in moir{\'e} minibands},
journal={Nature Communications},
year={2025},
month={Feb},
day={19},
volume={16},
number={1},
pages={1770},
issn={2041-1723},
doi={10.1038/s41467-025-57035-x},
url={https://doi.org/10.1038/s41467-025-57035-x}
}

@article{Hastings_2013,
  title = {Metaplectic anyons, Majorana zero modes, and their computational power},
  author = {Hastings, Matthew B. and Nayak, Chetan and Wang, Zhenghan},
  journal = {Phys. Rev. B},
  volume = {87},
  issue = {16},
  pages = {165421},
  numpages = {12},
  year = {2013},
  month = {Apr},
  publisher = {American Physical Society},
  doi = {10.1103/PhysRevB.87.165421},
  url = {https://link.aps.org/doi/10.1103/PhysRevB.87.165421}
}

@article{leakage_PhysRevA.97.032306,
  title = {Quantification and characterization of leakage errors},
  author = {Wood, Christopher J. and Gambetta, Jay M.},
  journal = {Phys. Rev. A},
  volume = {97},
  issue = {3},
  pages = {032306},
  numpages = {17},
  year = {2018},
  month = {Mar},
  publisher = {American Physical Society},
  doi = {10.1103/PhysRevA.97.032306},
  url = {https://link.aps.org/doi/10.1103/PhysRevA.97.032306}
}

@article{Alicea2011,
  doi = {10.1038/nphys1915},
  url = {https://doi.org/10.1038/nphys1915},
  year = {2011},
  month = feb,
  publisher = {Springer Science and Business Media {LLC}},
  volume = {7},
  number = {5},
  pages = {412--417},
  author = {Jason Alicea and Yuval Oreg and Gil Refael and Felix von Oppen and Matthew P. A. Fisher},
  title = {Non-Abelian statistics and topological quantum information processing in 1D wire networks},
  journal = {Nature Physics}
}

@article{EIGHT_QDOT_LADDER_hsiao2023exciton,
  title = {Exciton Transport in a Germanium Quantum Dot Ladder},
  author = {Hsiao, T.-K. and Cova Fari\~na, P. and Oosterhout, S. D. and Jirovec, D. and Zhang, X. and van Diepen, C. J. and Lawrie, W. I. L. and Wang, C.-A. and Sammak, A. and Scappucci, G. and Veldhorst, M. and Demler, E. and Vandersypen, L. M. K.},
  journal = {Phys. Rev. X},
  volume = {14},
  issue = {1},
  pages = {011048},
  numpages = {17},
  year = {2024},
  month = {Mar},
  publisher = {American Physical Society},
  doi = {10.1103/PhysRevX.14.011048},
  url = {https://link.aps.org/doi/10.1103/PhysRevX.14.011048}
}

@article{TWO_DOT_KITAEV_EXP_2023realization,
  title={Realization of a minimal Kitaev chain in coupled quantum dots},
  author={Dvir, Tom and Wang, Guanzhong and van Loo, Nick and Liu, Chun-Xiao and Mazur, Grzegorz P and Bordin, Alberto and Ten Haaf, Sebastiaan LD and Wang, Ji-Yin and van Driel, David and Zatelli, Francesco and others},
  journal={Nature},
  volume={614},
  number={7948},
  pages={445--450},
  year={2023},
  publisher={Nature Publishing Group UK London}
}

@article{THREE_DOT_KITAEV_EXP_bordin2023crossed,
  title = {Crossed Andreev Reflection and Elastic Cotunneling in Three Quantum Dots Coupled by Superconductors},
  author = {Bordin, Alberto and Li, Xiang and van Driel, David and Wolff, Jan Cornelis and Wang, Qingzhen and ten Haaf, Sebastiaan L. D. and Wang, Guanzhong and van Loo, Nick and Kouwenhoven, Leo P. and Dvir, Tom},
  journal = {Phys. Rev. Lett.},
  volume = {132},
  issue = {5},
  pages = {056602},
  numpages = {6},
  year = {2024},
  month = {Feb},
  publisher = {American Physical Society},
  doi = {10.1103/PhysRevLett.132.056602},
  url = {https://link.aps.org/doi/10.1103/PhysRevLett.132.056602}
}

@article{CA_SOC_SPIELMAN1_lin2011spin,
  title={Spin--orbit-coupled Bose--Einstein condensates},
  author={Lin, Y-J and Jim{\'e}nez-Garc{\'\i}a, K and Spielman, Ian B},
  journal={Nature},
  volume={471},
  number={7336},
  pages={83--86},
  year={2011},
  publisher={Nature Publishing Group UK London}
}

@article{CA_SOC3_SPIELMAN2_valdes2021topological,
  title={Topological features without a lattice in Rashba spin-orbit coupled atoms},
  author={Vald{\'e}s-Curiel, A and Trypogeorgos, D and Liang, Q-Y and Anderson, RP and Spielman, IB},
  journal={Nature communications},
  volume={12},
  number={1},
  pages={593},
  year={2021},
  publisher={Nature Publishing Group UK London}
}

@article{CA_SOC2_galitski2013spin,
  title={Spin--orbit coupling in quantum gases},
  author={Galitski, Victor and Spielman, Ian B},
  journal={Nature},
  volume={494},
  number={7435},
  pages={49--54},
  year={2013},
  publisher={Nature Publishing Group UK London}
}

@article{CA_Ladder_li2013topological,
  title={Topological states in a ladder-like optical lattice containing ultracold atoms in higher orbital bands},
  author={Li, Xiaopeng and Zhao, Erhai and Vincent Liu, W},
  journal={Nature communications},
  volume={4},
  number={1},
  pages={1523},
  year={2013},
  publisher={Nature Publishing Group UK London}
}

@article{FQH_para_clarke2013exotic,
  title={Exotic non-Abelian anyons from conventional fractional quantum Hall states},
  author={Clarke, David J and Alicea, Jason and Shtengel, Kirill},
  journal={Nature communications},
  volume={4},
  number={1},
  pages={1348},
  year={2013},
  publisher={Nature Publishing Group UK London}
}

@article{FQH_para_PhysRevX.2.041002,
  title = {Fractionalizing Majorana Fermions: Non-Abelian Statistics on the Edges of Abelian Quantum Hall States},
  author = {Lindner, Netanel H. and Berg, Erez and Refael, Gil and Stern, Ady},
  journal = {Phys. Rev. X},
  volume = {2},
  issue = {4},
  pages = {041002},
  numpages = {22},
  year = {2012},
  month = {Oct},
  publisher = {American Physical Society},
  doi = {10.1103/PhysRevX.2.041002},
  url = {https://link.aps.org/doi/10.1103/PhysRevX.2.041002}
}

@article{Fleckenstein2019,
  title = {{${\mathbb{Z}}_{4}$} parafermions in Weakly Interacting Superconducting Constrictions at the Helical Edge of Quantum Spin Hall Insulators},
  author = {Fleckenstein, C. and Ziani, N. Traverso and Trauzettel, B.},
  journal = {Phys. Rev. Lett.},
  volume = {122},
  issue = {6},
  pages = {066801},
  numpages = {6},
  year = {2019},
  month = {Feb},
  publisher = {American Physical Society},
  doi = {10.1103/PhysRevLett.122.066801},
  url = {https://link.aps.org/doi/10.1103/PhysRevLett.122.066801}
}

@misc{mccann2026noninteracting,
      title={Noninteracting tight-binding models for Fock parafermions}, 
      author={Edward McCann},
      year={2026},
      eprint={2510.07029},
      archivePrefix={arXiv},
      primaryClass={cond-mat.mes-hall},
      url={https://arxiv.org/abs/2510.07029}, 
}

@article{Knapp_2016,
  title = {The Nature and Correction of Diabatic Errors in Anyon Braiding},
  author = {Knapp, Christina and Zaletel, Michael and Liu, Dong E. and Cheng, Meng and Bonderson, Parsa and Nayak, Chetan},
  journal = {Phys. Rev. X},
  volume = {6},
  issue = {4},
  pages = {041003},
  numpages = {38},
  year = {2016},
  month = {Oct},
  publisher = {American Physical Society},
  doi = {10.1103/PhysRevX.6.041003},
  url = {https://link.aps.org/doi/10.1103/PhysRevX.6.041003}
}

@Article{Mills2019,
author={Mills, A. R.
and Zajac, D. M.
and Gullans, M. J.
and Schupp, F. J.
and Hazard, T. M.
and Petta, J. R.},
title={Shuttling a single charge across a one-dimensional array of silicon quantum dots},
journal={Nature Communications},
year={2019},
month={Mar},
day={05},
volume={10},
number={1},
pages={1063},
issn={2041-1723},
doi={10.1038/s41467-019-08970-z},
url={https://doi.org/10.1038/s41467-019-08970-z}
}

@article{Baart_2016,
    author = {Baart, T. A. and Jovanovic, N. and Reichl, C. and Wegscheider, W. and Vandersypen, L. M. K.},
    title = {Nanosecond-timescale spin transfer using individual electrons in a quadruple-quantum-dot device},
    journal = {Applied Physics Letters},
    volume = {109},
    number = {4},
    pages = {043101},
    year = {2016},
    month = {07},
    issn = {0003-6951},
    doi = {10.1063/1.4959183},
    url = {https://doi.org/10.1063/1.4959183},
}

@Article{DeSmet2025,
author={De Smet, Maxim
and Matsumoto, Yuta
and Zwerver, Anne-Marije J.
and Tryputen, Larysa
and de Snoo, Sander L.
and Amitonov, Sergey V.
and Katiraee-Far, Sam R.
and Sammak, Amir
and Samkharadze, Nodar
and G{\"u}l, {\"O}nder
and Wasserman, Rick N. M.
and Greplov{\'a}, Eli{\v{s}}ka
and Rimbach-Russ, Maximilian
and Scappucci, Giordano
and Vandersypen, Lieven M. K.},
title={High-fidelity single-spin shuttling in silicon},
journal={Nature Nanotechnology},
year={2025},
month={Jul},
day={01},
volume={20},
number={7},
pages={866-872},
issn={1748-3395},
doi={10.1038/s41565-025-01920-5},
url={https://doi.org/10.1038/s41565-025-01920-5}
}

@article{Das_Sarma_RevModPhys.80.1083,
  title = {Non-Abelian anyons and topological quantum computation},
  author = {Nayak, Chetan and Simon, Steven H. and Stern, Ady and Freedman, Michael and Das Sarma, Sankar},
  journal = {Rev. Mod. Phys.},
  volume = {80},
  issue = {3},
  pages = {1083--1159},
  numpages = {0},
  year = {2008},
  month = {Sep},
  publisher = {American Physical Society},
  doi = {10.1103/RevModPhys.80.1083},
  url = {https://link.aps.org/doi/10.1103/RevModPhys.80.1083}
}

@article{mzm_shuttle_truong_1,
  title = {Optimizing the transport of Majorana zero modes in one-dimensional topological superconductors},
  author = {Truong, Bill P. and Agarwal, Kartiek and Pereg-Barnea, T.},
  journal = {Phys. Rev. B},
  volume = {107},
  issue = {10},
  pages = {104516},
  numpages = {13},
  year = {2023},
  month = {Mar},
  publisher = {American Physical Society},
  doi = {10.1103/PhysRevB.107.104516},
  url = {https://link.aps.org/doi/10.1103/PhysRevB.107.104516}
}

@article{mzm_shuttle_truong_2,
  title = {Shuttling Majorana zero modes in disordered and noisy topological superconductors},
  author = {Truong, Bill P. and Agarwal, Kartiek and Pereg-Barnea, T.},
  journal = {Phys. Rev. B},
  volume = {113},
  issue = {6},
  pages = {064506},
  numpages = {17},
  year = {2026},
  month = {Feb},
  publisher = {American Physical Society},
  doi = {10.1103/tswn-kxhx},
  url = {https://link.aps.org/doi/10.1103/tswn-kxhx}
}

@article{Osvath_2024,
  title = {Electronic ladder model harboring {${\mathbb{Z}}_{4}$} parafermions},
  author = {Osv\'ath, Botond and Barcza, Gergely and Legeza, \"Ors and D\'ora, Bal\'azs and Oroszl\'any, L\'aszl\'o},
  journal = {Phys. Rev. B},
  volume = {110},
  issue = {8},
  pages = {085304},
  numpages = {11},
  year = {2024},
  month = {Aug},
  publisher = {American Physical Society},
  doi = {10.1103/PhysRevB.110.085304},
  url = {https://link.aps.org/doi/10.1103/PhysRevB.110.085304}
}
\end{document}